\documentclass[preprint]{aastex}

\shorttitle{Close binary stars.~IV}
\shortauthors{Lu, Rucinski \& Og{\l}oza}

\begin{document}

\title{Radial Velocity Studies of Close Binary 
Stars.~IV\footnote{Based on the data obtained at the David Dunlap 
Observatory, University of Toronto.}}

\author{Wenxian Lu\footnote{Current address:
  NASA/GSFC, Mailstop 664, 8800 Greenbelt Road, Greenbelt, MD 20771}  
~~and Slavek M. Rucinski}
\affil{David Dunlap Observatory, University of Toronto\\
P.O.Box 360, Richmond Hill, Ontario, Canada L4C 4Y6}
\email{lu,rucinski@astro.utoronto.ca}

\and

\author{Waldemar Og{\l}oza}
\affil{Mt. Suhora Observatory of the Pedagogical University\\
ul.~Podchora\.{z}ych 2, 30-084 Cracow, Poland\\ and
N.Copernicus Astronomical Center\\
ul.~Bartycka 18, 00-716 Warszawa, Poland}
\email{ogloza@wsp.krakow.pl}


\begin{abstract}

Radial-velocity measurements and sine-curve fits to the orbital 
velocity variations are presented for the fourth set of ten close binary 
systems: 44~Boo, FI~Boo, V2150~Cyg, V899~Her, EX~Leo, VZ~Lib,
SW~Lyn, V2377~Oph, Anon Psc (GSC~8-324), HT~Vir. All systems are 
double-lined spectroscopic binaries with only two of them not being
contact systems (SW~Lyn and GSC~8-324) and with five
(FI~Boo, V2150~Cyg, V899~Her, EX~Leo, V2377~Oph) being the recent
photometric 
discoveries of the Hipparcos satellite project. Five of the binaries
are triple-lined systems (44~Boo, V899~Her, VZ~Lib, SW~Lyn, HT~Vir). 
Three (or possibly four) companions in the triple-lined systems
show radial-velocity changes during the span of our observations
suggesting that these are in fact quadruple systems. 
Several of the studied systems are prime candidates for combined 
light and radial-velocity synthesis solutions.
\end{abstract}


\section{INTRODUCTION}
\label{sec1}

This paper is a continuation of the radial-velocity studies of close 
binary stars, \citet[Paper I]{ddo1}, \citet[Paper II]{ddo2} and
\citet[Paper III]{ddo3}.
The main goals and motivations are described in these papers.
In short, we attempt to obtain radial-velocity data for
close binary systems which are accessible to 1.8-m class 
telescopes at medium spectral resolution of about R = 10,000 -- 15,000. 
Selection of the objects is quasi-random in the sense that we have
started with short period (mostly contact) binaries and we
attempt to slowly progress to longer periods as the project
continues.
We publish the results in groups of ten systems as soon as reasonable
orbital elements are obtained from measurements evenly distributed 
in orbital phases. 
  
This paper is structured in the same way as Papers~I -- III in that 
most of the data for the observed binaries are in 
two tables with the radial-velocity measurements 
(Table~\ref{tab1}) and their sine-curve solutions (Table~\ref{tab2}).
Section~\ref{sec2} of the paper contains brief summaries of 
previous studies for individual systems. The special feature
of this paper which distinguishes it from the previous three papers
is a discussion of close spectroscopic/visual companions 
to five close systems in Section~\ref{sec3}. We found many
similarities in these triple-lined 
systems and decided to publish them together
in one installment of our series. Table~\ref{tab3} lists the
radial-velocity data for the companions which we always call
the ``third components'' of the systems. In this spirit, we will
discuss their luminosities in relation to those of the
close binaries, $L_3/L_{12}$. We utilize the conventional
naming of such components in visual systems with A signifying the
brighter component in a close visual pair (HT Vir turns out
to be an exception).

The observations reported in this paper have been collected between
February 1997 and October 2000; the ranges of dates for
individual systems can be found in Table~\ref{tab1}.
Eight systems discussed in this paper have been observed for
radial-velocity variations for the first time. 44~Boo and SW~Lyn
have been observed before; for both, we are providing much 
improved radial-velocity orbits.

We derive the radial-velocity data using the {\it broadening function\/}
(BF) approach of the linear Singular Value Decomposition (SVD),
as described in \citet{ruc99}\footnote{Practical
advice and detailed suggestions are available also in 
http://www.astro.utoronto.ca/$\sim$rucinski/.}.
The two peaks of the BF are fitted by Gaussians giving radial
velocities of the components. Then, variations of the radial
velocities are represented by sine curves giving the center-of-mass
velocity, $V_0$, the two semi-amplitudes, $K_1$ and $K_2$, and the phase
of the primary eclipse, $T_0$. Since our initial goals
were the values of $V_0$ and of the mass-ratio $q = K_1/K_2$,
this approach was deemed adequate when we started the program. 
We are aware that the
amplitudes and hence the masses may be biased by the 
Gaussian approximation, but we are unable to do the full
modeling (as was done for AW~UMa, AH~Vir and W~UMa: 
\citet{ruc92,lu93,ruc93}),
without knowledge of the degree-of-contact and of the orbital
inclination which are derivable from light curves. Thus, we
continue using the Gaussians recognizing that one day our
values of the semi-amplitudes may have to be corrected for
systematic inaccuracies of our measurements.

The data in Table~\ref{tab2} are organized in the same manner as 
before. In addition to the parameters of spectroscopic orbits,
the table provides information about the relation between 
the spectroscopically observed epoch of the primary-eclipse T$_0$ 
and the recent photometric determinations in the form of the (O--C) 
deviations for the number of elapsed periods E. It also contains
our new spectral classifications of the program objects (unless
taken in parentheses when we use other spectral type estimates).

For further technical details and conventions used in the paper, please
refer to Papers I -- III of this series.

\placetable{tab1}

\placetable{tab2}

\section{RESULTS FOR INDIVIDUAL SYSTEMS}
\label{sec2}

\subsection{44 Boo B}

With the large parallax of $p=78.4 \pm 1.0$ milli-arcsec (mas)
\citep[HIP]{hip},
44~Boo~B\footnote{The star 44~Boo has another name, 
i~Boo. Frequently, these are combined, which is incorrect and
makes electronic searches difficult. 
We use only the former name in this paper.} 
is the nearest contact binary and one of the nearest
close binaries. Because of its
brightness, it has been one of the most frequently photometrically 
observed variable stars in the sky. 44~Boo~B is 
the fainter component of a relatively close 
(currently 1.7 arcsec) visual double which is
attracting vigorous observational activity in the fields of
stellar interferometry and speckle-interferometry.
We do not relate to these studies
because our radial-velocity data have been obtained in a 
short interval of only 10 days so that they contribute only
one radial-velocity data point 
of the long-period (225 year), visual orbit.

Several studies attempted to determine a radial-velocity orbit
for 44~Boo~B. These efforts are summarized in \citet{hil89}.
Here we present results far superior to any previous attempts.
The success is mostly due to extraction of the component velocities
using the broadening function approach
(see Section~\ref{sec3}) which permitted excellent separation
of the three spectral components in the system.
In particular, we provide an accurate estimate of the mass 
ratio for 44~Boo~B, $q_{sp} = 0.487 \pm 0.006$, which should 
be of much help
in future photometric studies of the contact system. The
radial-velocity orbit is shown graphically in Figure~\ref{fig1}.

Since the discovery \citep{sch26}, the presence of the brighter
companion created problems in studies of the close pair. The
most extensive study of 44~Boo system by \citet{hil89} discussed 
efforts of de-coupling of the spectral and photometric signatures 
of the visual pair, but even such basic data as
spectral types or color indices were difficult to derive.
The re-analysis of the Hipparcos data for close visual binaries 
\citep{fab00}, gave reliable magnitudes and
color indices for both components, $V^A=5.28$, $(B-V)^A = 0.65$
and $V^B=6.12$, $(B-V)^B=0.94$; the magnitudes are the
average ones so that they do not take into account the variability of 
component B. The color index for component A 
agrees with the estimate of the spectral type by \citet{hil89}
of G1V while the color index for component B points at
a much later spectral type than was considered before, 
around K2V, in accordance with the well
known high magnetic activity of the contact system. 

The magnitude
difference at maximum of 44~Boo~B relative to component A
was estimated by \citet{hil89}
as $\Delta V =0.63$, so that $V^B(max)=5.91$. For this
brightness, the absolute magnitude predicted from the parallax is 
$M_V^B = 5.38 \pm 0.04$, whereas the prediction of the 
$M_V(\log P, B-V)$ calibration \citep[RD97]{RD97}
is $M_V^B(cal) = 5.50$. These estimates would agree at $M_V^B = 5.38$ 
if we use $(B-V)=0.90$ in the
calibration, assuming that the maximum light $(B-V)$ is slightly
smaller (bluer) than the average value given by \citet{fab00}. We note
that 44~Boo~B, in spite of its proximity, was used in the first
$M_V(\log P, B-V)$ with some reservations \citep{ruc94} because
its color index was poorly know at that time. 
It is gratifying to see now a much
better agreement of the directly determined $M_V$
for 44~Boo~B with the calibration.

\placefigure{fig1}

\subsection{FI Boo}
FI Boo is a new discovery from the Hipparcos satellite mission.
\citet{due97} included it in the list of possible contact binaries.
The period reported in the Hipparcos Catalogue is equal to
one half of the orbital period that we find
and the zero epoch $T_0$ was given in the Catalogue
for the maximum light. 
With our spectroscopic results, the contact system 
is of the W-type, with the less
massive component eclipsed at the eclipse given by $T_0$, which
has been selected to be one quarter of the orbit before the maximum
given in HIP. 

The radial-velocity orbit for FI Boo is very well determined
(Figure~\ref{fig1}).
The small value of $(M_1 + M_2) \sin^3i = 0.343 \pm 0.010 \, M_\odot$ 
is consistent with the small photometric amplitude of 
about 0.15 mag.\ in suggesting a low orbital inclination. 

Assuming $V_{max} = 9.57$ on the basis of the 
average $V$ in \citet[TYC2]{tyc2} and the amplitude in HIP,
with the parallax of $p =9.52 \pm 2.10$ mas, we obtain
$M_V = 4.41 \pm 0.49$. The RD97 calibration gives $M_V(cal)=4.17$.

\subsection{V2150 Cyg}

V2150 Cyg has been discovered by the Hipparcos satellite mission. 
It shows a small photometric amplitude of 0.12 mag.\ 
suggesting a low orbital
inclination. This is in accord with the low value of
$(M_1 + M_2) \sin^3i = 1.376 \pm 0.018 \, M_\odot$ for an early 
spectral type of the binary which  suggests a rather 
massive system. \citet{abt81} estimated the spectral type to be
A5V; our independent estimate is A6V. The orbital period is 0.591
days and the HIP light curve looks like of a genuine contact binary,
which is a rare occurrence among such early-type systems. 

Although the system is a visual double (ADS 14835), its orientation
relative to the spectrograph slit was such that we were able
to avoid the faint ($\Delta m = 3.4$)
companion at the separation of 3.7 arcsec.

Our radial-velocity orbit is very well determined (Figure~\ref{fig1}). 
The contact system consists of two very similar components
with $q=0.802 \pm 0.006$ with a more massive component
eclipsed at the primary minimum (type A).
The only previous radial-velocity measurements of 
V2150~Cyg, in a survey by \citet{gre99}, showed scatter 
among three measurements indicating radial-velocity variability:
$V = 31.4 \pm 16.8$ km~s$^{-1}$. 
The nominal accuracy of this survey was 3.0 km~s$^{-1}$
on the basis of spectra with resolution about 8 times lower than
ours. 

In spite of a good solution of the radial-velocity orbit, we found
one disturbing feature of the solution for V2150~Cyg: The
radial-velocity amplitudes depend on the spectral type of the 
template star used to derive the broadening functions. This is 
unusual because for most W~UMa-type systems we have not seen 
any dependence of the amplitudes on the spectral type. Normally, 
with some spectral-type mismatch of the template, the broadening function
could change its intensity scale and its quality of determination, but the
radial-velocity amplitudes would stay constant. We selected several
sharp-line standards from the list of Dr.\ Frank Fekel
(private communication). Over the range of the template spectral types
within the A-type, we see a change of up to 7\% in both semi-amplitudes 
$K_i$, with larger amplitudes obtained with templates
of earlier spectral types. While the mass-ratio remained 
perfectly constant, the systematic uncertainty in the
amplitudes would obviously affect the derived values of masses. We
have no explanation of the effect, but note that V2150~Cyg is
of the earliest spectral type among binary systems analyzed by us
so far.

The system of V2150~Cyg is important because it may provide an
crucial extension of the absolute magnitude calibration
$M_V(\log P, B-V)$ toward early spectral
types. With $V_{max}=8.06$ estimated from the average data in TYC2
and the amplitude, and with the parallax
$p=4.7 \pm 1.6$ mas, we obtain 
$M_V=1.43 \pm 0.74$ while the RD97 calibration,
pushed to a color index as blue as $(B-V)=0.25$,
gives $M_V(cal)=1.89$. A big question is
obviously the amount of reddening for this relatively distant (about 210 pc)
system. The color index, when compared with the
spectral type A5/6~V, suggests a moderate
reddening of about $E_{B-V} = 0.07$. In this case, the absolute
magnitude derived from the parallax is
$M_V=1.21 \pm 0.74$, while $M_V(cal)=1.67$.

\subsection{V899 Her B}

V899 Her is the next Hipparcos discovery. Although it is classified
in the Hipparcos catalogue as an EB light-curve system
(with unequal minima), it seems to
be a genuine contact binary with a low-amplitude
(0.14 mag.) EW light curve and 
practically equally deep eclipses. Apparently, the
light curve classification was driven by one deviating point,
possibly a photometric error. We
have kept the Hipparcos ephemeris which leads to an A-type
contact system.

We discovered that
V899 Her is a spectroscopically triple system in which 
the contact binary is the fainter component, so that
we designate the contact system as component B. The broadening function
was an indispensable tool in separating the three spectral
components; without it, the system would be probably
unsolvable (see Section~\ref{sec3}).
The presence of the bright companion explains the inconsistency
between large radial-velocity amplitudes of the components
of V899~Her~B, leading to a large value of 
$(M_1 + M_2) \sin^3i = 2.33 \pm 0.08 \, M_\odot$
(and thus an indication of the orbital inclination close to $90^\circ$)
and the small photometric amplitude of the W~UMa-type
light curve which appears to be ``diluted'' in the total systemic light.

The spectral type that we estimated from our low-resolution
classification spectra is F5V, with indications of a
contribution from the fainter G-type component.
Because of the dominance of the third component,
there is no point to apply the RD97 calibration to this
case. The maximum brightness and the color index,
$V_{max}=7.87$, $(B-V)=0.48$ almost certainly reflect the
properties of the third star rather than that of the close binary.

Because of the presence of the bright companion, the radial-velocity
orbit of V899~Her~B is not as well defined as for other
close binaries (Figure~\ref{fig1}). The bright
companion A is itself a radial-velocity variable because we 
noticed well defined variations within the 250 day span of our 
observations (see Section~\ref{sec3}). We do not see these variations 
in the systemic velocity of the close pair, so 
we assume that the visual companion is itself an independent
spectroscopic binary. 

\placefigure{fig2}

\subsection{EX Leo}

EX Leo has been found by Hipparcos. Our solution (Figure~\ref{fig2})
describes a rather 
typical contact binary of the A-type with a small mass-ratio,
$q =0.199 \pm 0.036$. 
Our estimate of the spectral type, F6~V,  
agrees very well with $(B-V) = 0.53$ from TYC2. 
The light curve has an amplitude of about
0.25 mag.\ and is quite well defined, but perhaps a bit sparsely
covered by the HIP data. 

The parallax $p=9.84 \pm 1.11$ mas with $V_{max}=8.17$ leads to
$M_V=3.13 \pm 0.26$ so that it agrees with the RD97 calibration,
$M_V(cal)=3.45$.

\subsection{VZ Lib A}

VZ Lib has been known as a contact binary since
the work of \citet{tse54}. 
\citet{cla81} published the last currently available 
photometric study of the system. 

We have detected a spectroscopic companion to VZ~Lib, about five
times fainter than the contact binary. We discuss
it in Section~\ref{sec3}. Here we note that this component may have a 
slowly variable radial velocity, but these variations are not reflected
in the systemic velocity of the close binary. 

It is impossible to derive the correct $V_{max}$ for the contact
binary value because of the poorly known 
contribution of the third component to
the total systemic light. The observed combined brightness,
$V_{max} = 10.36$, $(B-V)=0.61$ \citep{cla81}
is accord with our estimate of the spectral type,
G0 (composite). These data, together with the HIP parallax
$p=4.92 \pm 1.96$ mas imply $M_V=3.59 \pm 0.88$ for the whole 
system while the RD97 calibration for the contact system
predicts $M_V(cal)=3.94$ which is in agreement with some 
contribution of the third star to the systemic brightness.

Because of the faintness of the system, presence of the
third component, and difficulties of measuring velocities
of the close binary, our radial-velocity solution is relatively 
poor (Figure~\ref{fig2}).
The system appears to be a contact binary 
of the A-type with a moderately
small mass-ratio of $q=0.24 \pm 0.07$. The orbit is probably
oriented close to edge-on because the sum of masses 
is relatively large, $(M_1 + M_2) \sin^3i = 1.70 \pm 0.12 \, M_\odot$. 

We note that our spectroscopically derived moment of the primary
eclipse deviates by about 2 hours from the ephemeris of
\citet{cla81}, either due to period changes or accumulated
errors in the period over the long time since the last 
photometric observations.

\subsection{SW Lyn A}

The binary was discovered as a variable star by \citet{hof49}. Major
photometric studies were done by \citet{vet68,vet77}. Recently,
\citet{ogl98} re-discussed the extant data; the discussion was
to a large extent guided by the spectroscopic results of \citet{vet77}
in that a mass-ratio was assumed to be around $q \simeq 0.35$, in spite
of photometric solutions which tended to converge
to $q \simeq 0.5 - 0.6$.
In fact, our spectroscopic results do give $q_{sp} = 0.52 \pm 0.03$.

From the spectroscopic point of view, 
SW Lyn is quite complex: It appears to be a very
close, short-period Algol system with a faint, but
detectable secondary. We also clearly
see a third component, about three times fainter
than the close binary, which has not previously been detected 
spectroscopically (see Section~\ref{sec3}). 
Possibly, it is the same star which is producing 
the 2128 day periodicity in the eclipse timing \citep{ogl98}.

Our radial-velocity orbit is entirely different from that of
\citet{vet77} which was obtained at a very low spectral resolution. Thus,
we do not confirm any manifestations of apparent eccentricity which
were explained by gas streams. In view of our results, 
the detailed discussion of the
physical properties of the system by \citet{ogl98} may have to
be revised. We did experience difficulties
in measuring velocities of the secondary component whose
signal in the broadening function is very weak in comparison
with the third star and the primary component (see Section~\ref{sec3}).
These difficulties were especially severe in the
second half of the orbit. Thus, in order not to affect the systemic 
velocity $V_0$ in our combined orbital solution which takes
into account the velocities of both stars, we 
assigned weights of one-half and one-quarter 
to all measurements of the secondary within
the first and second halves of the orbit, respectively
(Figure~\ref{fig2}).

Our observations were obtained in two groups (16 and 55 observations) 
separated by one year which -- in view of the presence of the third
star -- poses a question whether the observations of 
the close binary could be combined in one solution. 
Accordingly, we obtained one solution based
on all observations and one based only on the 55 
observations of the second season (the phase distribution
for the first season did not permit a separate solution). 
The results do not differ
significantly, indicating that the center of mass of the close binary
probably did not move much between the seasons. However, the
conjunction times came out different, but within the combined errors
of both solutions. For reference, we give here the results of the
solution based only on the second season: $V_0=+31.11(1.26)$,
$K_1=116.09(1.52)$, $K_2=222.87(3.60)$, $T_0=2,451,400.1752(22)$.

The photometric properties of SW~Lyn are rather poorly known.
\citet{vet68} estimated $V_{max}= 9.2$ and $(B-V)=0.38$. The
color index corresponds to the spectral type of F2~V. However,
TYC2 suggests the mean $(B-V)=0.23$. 
With its period of 0.644 days, SW~Lyn is of interest
to studies of the short period Algol systems.

Our spectroscopic value of $T_0$ agrees very well with the
recent photometric timing of \citet{ogl00}.

\subsection{V2377 Oph}

The binary has been discovered by the Hipparcos satellite. It appears to
be a fairly uncomplicated W-type contact binary and our solution
is very well defined (Figure~\ref{fig2}). The small value of 
$(M_1 + M_2) \sin^3i = 0.487 \pm 0.009 \, M_\odot$ is in accordance
with the small amplitude of photometric variations (0.04 mag.), both
resulting from a low inclination angle.

Our spectral type G0/1~V is in slight disagreement with the TYC2 
color index of $(B-V)=0.67$ so that some reddening 
of about $E_{B-V} \simeq 0.07$ is possible. The system is relatively
distant (about 100 pc) and has
a parallax $p=10.09 \pm 1.22$ mas which implies $M_V = 3.53 \pm 0.27$
for the case of no reddening, while 
the RD97 calibration predicts $M_V(cal)=3.79$.
For $E_{B-V} = 0.07$ the absolute magnitudes would be
$M_V=3.32$ and $M_V(cal)=3.58$, respectively.

\placefigure{fig3}

\subsection{Anon Psc = GSC 8--324}

This very interesting close binary has been discovered recently by
\citet{rob99}. It does not yet have a variable star name, so we are 
using here an abbreviated one, GSC~8--324\footnote{The name GSC 8--324 is
in fact incorrect which may complicate electronic searches.
It should be GSC 00008--00324. 
The notation used by \citet{rob99}, who used a different
numbers of leading zeros, is also incorrect.}.

The system appears to be a very close pair of two detached 
K-type dwarfs on tight orbit with the orbital period of $P = 0.3086$
days. We included the system in our program soon after learning about
its discovery, led by an expectation that this is a new case of
V361~Lyr-like binary which is evolving rapidly into contact
\citep{hil97}, but some 4 magnitudes brighter, therefore easier for
detailed studies. 
Recent correspondence with Dr.\ Robb (private communication)
indicated that the spots on the surface of GSC~8--324 have moved 
between the seasons so that the system is {\it not\/} 
similar to V361~Lyr where the high stability of the
accretion region strongly suggests a stable flow of matter between
components. However, because the system is so very tight and yet
detached -- as judged by short duration of eclipses -- and also
relatively nearby ($14 \pm 8$ pc, see below), 
it is one of the most interesting 
among recently discovered close binaries.

Because of the late type of the components, K4-5V (see references in
\citet{rob99}) we feared that the magnesium triplet lines would not
be as well suited for radial velocity observations as for F--G
stars. For that
reason we made a change in our observational setup and, 
in addition to the observing the same region 5185 \AA\ as 
for other stars, we also
obtained observations in the region centered at 5303 \AA. The results
of independent solutions were identical so that we made a combined
solution for both spectral regions. However, we list the two series of
observations separately in Table~\ref{tab1}. The phased observations
of the system are shown in Figure~\ref{fig3}.

The radial-velocity data for the primary component are well defined
for both spectral regions. In contrast, the secondary-component
peaks in the broadening function are weak and the radial-velocity data 
are correspondingly poorer. Yet, the mass-ratio is very 
well determined, $q_{sp} = 0.702 \pm 0.014$. 
The value of $(M_1 + M_2) \sin^3i = 1.04 \pm 0.02 \, M_\odot$ 
together with the presence of deep eclipses suggests a possibility 
of a reliable
combined solution for absolute elements of the system. If not for
the presence of large photospheric spots, the system could serve as
one of the lower main sequence calibrators.

The epoch of the superior conjunction (the primary eclipse) indicates
that the orbital period could be slightly adjusted relative to the
value given by \citet{rob99}, but we leave this matter open as we
cannot exclude period changes in such a short-period system.

The system was included in the Tycho-1 catalogue (HIP). The
parallax has a large error, but does suggest a relatively nearby
system, $p=72 \pm 40$ mas. With the data
in \citet{rob99}, $M_V = 9.9 \pm 1.2$. The system 
has a large and relatively well-defined proper motion, 
$\mu_{RA}\cos{\delta} = -108.2 \pm 2.2$ mas/yr and 
$\mu_{dec} = -201.1 \pm 2.1$ mas/yr (TYC2), undoubtedly due to
the proximity of the system.

\subsection{HT Vir B}

HT Vir belongs to a very close visual binary, 
a situation somewhat similar to that of 44~Boo, except that the
separation is smaller, below one arcsec, and the brightnesses
of the visual components are similar, with the contact
binary being fainter only during the eclipses. 
We retain the designation of B for the contact system, 
for consistency with the previous investigations. 
The close visual binary
is currently the subject of numerous interferometric and 
speckle-interferometry studies.

The re-analysis of the Hipparcos data for visual binaries 
\citep{fab00}, gave reliable magnitudes and
color indices for both components at the time that the
separation was only 0.56 arcsec: $V^A=7.80$, $(B-V)^A = 0.64$
and $V^B=8.30$ (average magnitude), $(B-V)^B=0.56$. 
Our combined spectral type, with a strong contribution of
the third component, is F8V; this does not very well 
agree with its color index which suggests a later-type star.

HT Vir was discovered by \citet{wal84} and then studied photometrically
by \citet{wal85}. The photometric solution suggested that the third
component provides about as much light as the eclipsing
system at its light maxima. 
Our broadening function results confirm the strong
presence of the third component; we discuss this further in 
Section~\ref{sec3}. We found that the third component is 
in fact a radial-velocity variable with a period of 32.45 
days. Thus the system is really a quadruple one, with lines
of three systems visible in the spectra.
We give a preliminary solution for the single-lined (SB1) 
system HT~Vir~A in Section~\ref{sec3} and in Table~\ref{tab4}. 

Because we have been able to isolate the signatures of HT~Vir~A
from those of the contact binary, the radial-velocity 
solution for HT~Vir~B is very well defined (Figure~\ref{fig3}). 
We note that
the binary is of the A-type, yet has a surprisingly large
mass ratio for such a system, $q=0.812 \pm 0.008$. It is interesting 
to note that, although our determination of $q$ 
for HT~Vir~B is the first one,
\citet{wal85} postulated a value not far from unity on the basis
of the ratio of radii derived from their photometric analysis.
The large mass ratio would be in accord with the relatively large
amplitude of light variations of HT~Vir. \citet{wal85} estimated
that $L_3/(L_{12}+L_3) \simeq 0.44$ and thus
$L_3/L_{12} \simeq 0.79$ at light maxima of the close
binary; HT~Vir~A would be then the slightly fainter of the
two visual components. The light variation amplitude observed by
\citet{wal85} of about 0.42 mag., when corrected for the contribution 
of HT~Vir~A to the total systemic brightness would be then 
about 0.92 mag.
Such a large amplitude can be generated only by a contact system
with the orbit seen edge-on, but also with a mass ratio close to
unity. Integrations of the individual spectral features in our
broadening functions (Section~\ref{sec3}) suggest 
$L_3/L_{12} \simeq 0.52 \pm 0.05$ at the light maxima of the
contact system so that HT~Vir~A was 
observed by us to be fainter relative to the contact binary
than before; its spectral signature 
was always better defined in the spectra and in the
broadening functions because of its slow rotation.

The parallax of the system is the largest in this group after that
of 44~Boo, $p=15.39 \pm 2.72$ mas. With $V_{max}^B = 8.1$
and $(B-V)^B=0.56$, this implies $M_V^B = 4.0 \pm 0.4$. 
The RD97 calibration predicts $M_V^B(cal)=3.54$.  

Our spectroscopic determination of $T_0$ shows a relatively large
O--C deviation, a result of using the ephemeris of \citet{wal85}.
Similarly to VZ~Lib, HT~Vir has not been  observed for
eclipse timing for a long time.

\section{BROADENING FUNCTION APPROACH AND SPECTROSCOPIC COMPANIONS}
\label{sec3}

\subsection{Broadening functions}

The spectroscopic signatures of the visual/spectroscopic companions
are particularly well defined in broadening functions (BF)
which give projection of a system into the velocity space \citep{ruc99}. 
While the close binaries of our program have short periods and thus have
their rotation/revolution signatures spread over wide ranges of
radial velocities, the third components -- which rotate slowly
-- show sharp peaks in the BF. We can easily model these sharp peaks
by applying the BF derivation to other sharp-line stars.

Figure~\ref{fig4}
shows the BF's for the five triple systems, 44~Boo, V899~Her,
VZ~Lib, SW~Lyn and HT~Vir, all at phases close to orbital 
quadratures of the close pairs. In one case of HT~Vir, 
we show elements of the BF decomposition into the third component
and the binary itself (the last panel of Figure~\ref{fig4}). 
Of particular importance is that the third component can be
cleanly subtracted from the BF leaving a very well defined 
signature of the contact system. The success of the
decomposition is due primarily to the linear properties of
the BF derived through the SVD formalism \citep{ruc99},
in contrast to the cross-correlation function which is non-linear.
This permits subtraction of the 
third component from the BF which is simple and mathematically
correct. Then, through separate
integration of the sharp and broad components, we can
determine the brightness of the companion in relation to that
of the close binary, $L_3/L_{12}$. 
Thanks to the linear SVD--BF approach, 
we have been able to determine radial velocities for
components in close binary systems which would present 
totally insurmountable challenge if
handled with the cross-correlation function, 
such as V899~Her, VZ~Lib or SW~Lyn.

The linear deconvolution technique that we use to obtain the
broadening functions is not the only one available for determination
of accurate radial velocities from blended spectra showing lines
of several components. In recent years, a powerful technique called
TODCOR has been developed for close binaries 
by \citet{todcor1} and then extended
into the case of triple-lined systems by \citet{todcor3}; the latter
modification
has been already successfully applied in some difficult cases
of multi-lined systems
\citep{todcor3a}. We did not use this technique for the several
reasons. Not only that we feel more comfortable with a tool
developed by ourselves, but (1)~so far, TODCOR has 
not been demonstrated to work for very broad lines
of contact binaries, and (2)~our case of mixed very broad and narrow
spectral signatures is even more difficult and would require
even more extensive testing. We fear in particular that the
non-linear nature of the cross-correlation would complicate
derivation of relative brightnesses of components for systems with
components showing very different degrees of rotational broadening.

In three (V899 Her, SW Lyn, HT Vir), possibly four (VZ Lib), 
among the triple-lined systems, the third component 
appears to be a radial-velocity variable.  
We will comment on each system in the separate subsections below. 

\subsection{44 Boo A}

The observations lasted too short in the case of 44~Boo~A 
to note any long-term radial-velocity changes. However, 
since we wanted to obtain the best radial-velocity data for the
close binary 44~Boo~B, the presence of the bright companion
did require some additional precautions. In particular,
an attempt was made during the
observations to place the brighter component 44~Boo~A
as much outside the spectrograph slit as possible. The separation 
is currently about 1.7 arcsec so that it is comparable to
the width of our slit (1.8 arcsec), which cannot be rotated. 
Because the component A is about 1.8 times brighter than B
\citep{hil89} and its lines are sharp and very well defined, 
its signature is always visible in the BF 
(first panel of Figure~\ref{fig4}). By attempting to
place its image outside the slit, we partly succeeded in suppressing 
its contribution (to the level $L_3/L_{12} \simeq 0.4 - 0.7$
depending on the seeing, as measured in the broadening functions), 
but we did not eliminate it
totally achieving only modification of its relative
intensity in the BF. 
Its radial velocity derived from the BF is also
incorrect (by up to 12 km~s$^{-1}$) because the light was spilling 
over always from one side of our relatively wide slit. 
To derive an unbiased velocity of 44~Boo, 
we made three additional observations
of this component at the slit center and measured the
velocity of this component separately. Its average velocity for
the epoch JD(hel) = 2,450,945.996 
is $V^A = -34.93 \pm 0.64$. The detailed
data are given in Table~\ref{tab3}.

\subsection{V899 Her A}

There exist no earlier reports of the 
presence of a third component in this star
which has been only recently discovered as a variable star.  
As we can see in the second top panel of Figure~\ref{fig4}, 
the third component dominates the broadening function of V899~Her,
so we call it the component A. The ratio of brightnesses at phases
close to orbital quadratures is $L_3/L_{12}=1.5 \pm 0.1$.
The spectrum of the system, F5~V,
is also dominated by this component and we see only
weak signatures of the broadened G-type spectrum in the BF.
Only thanks to the linearity of the broadening function
determination we have been able to study this interesting system.

We found that V899~Her~A slowly changes its radial velocity. 
These variations are not reflected in the systemic velocity of
the close pair, so that the companion A itself is probably a
wide spectroscopic binary. The radial-velocity data
for V899~Her~A are shown in Figure~\ref{fig5}.
We had many observations for
this star so that we could group them into nightly average values. 
These in turn permitted us to obtain independent estimates of
measurement errors for individual observations, after all stages of
the combined standard and broadening-function processing
and component separation. For such a sharp-line star, 
the errors of the nightly averages are small at a level typically 
0.5 to 1.0 km~s$^{-1}$.

\subsection{VZ Lib B}

We have detected a spectroscopic companion of VZ~Lib. The broadening
function (the left lower panel in Figure~\ref{fig4}) 
indicates that in this case the brightness of the companion is
lower that that of the close binary, $L_3/L_{12}=0.20 \pm 0.04$.
Because the close
binary has a relatively small mass ratio, the BF peak of the third star
is usually merged with the broadening lobe of the primary, more massive
component of the W~UMa-type system of VZ~Lib.

The radial velocities of the third component show a nightly spread
of 2.6 to 4.3 km~s$^{-1}$, whereas errors for sharp
line stars are typically $\le 1$ km~s$^{-1}$. This may be an indication 
that the spectroscopic companion is itself a close binary
although we found that derivation of the velocities for all
three stars in this case was very difficult  because of the relative
faintness of the system and extensive blending of the components
in the broadening function. The variations of the third component
appear to be slow (Figure~\ref{fig5}), but may
result from grouped sampling of more rapid changes. The system
requires further radial-velocity observations to confirm that the
companion has variable radial velocity and that the system
is actually a quadruple one.

\subsection{SW Lyn B}

We have detected a third star in the system of SW~Lyn. It is well
visible in the broadening function in Figure~\ref{fig4} 
(middle lower panel). As mentioned in the description
of the close binary in Section~\ref{sec2},
the presence of the signal from the third star as well as the dominant
role of the massive and bright primary component of the
Algol system resulted in 
difficulties with the radial-velocity measurements of the secondary
star of the close system.

The third star in SW Lyn is moderately bright,
$L_3/L_{12}=0.33 \pm 0.05$ (for the contact system at
the orbital quadrature), and is a radial-velocity variable so that
the whole system is a quadruple one. The radial-velocity 
changes of the third component are
slow and certainly compatible with the orbital period of 2128 day 
which is noticeable in the eclipse timing for the close
pair \citep{ogl98}. However, we see no direct indication that this
is the same star. Unfortunately, contrary to V899~Her or VZ~Lib,
we have been unable to check if seasonal data would give the 
complementary systemic velocity variation
for the close pair; we could only solve the second
season or all the data together. As we described in Section~\ref{sec2}
the orbital solutions for the close pair based on all the
observations from both seasons and, separately, on 
the observations from the second season, gave identical sets of orbital
parameters (within the errors). This would suggest that
the spectroscopic companion and the eclipse-timing perturber are not
the same object. 

\subsection{HT Vir A}

HT Vir has been known as a close visual binary with a separation of less
than one arcsec \citep{wal84,wal85}. In terms of the overall 
properties, the system is somewhat similar to 44~Boo. 
\citet{wal85} determined that the component A
had brightness comparable to that of the close binary
($L_3/L_{12} \simeq 0.79$) and was slightly fainter of the two even
at light maxima of the contact system.
By integration of the separate
components of the broadening function (upper right panel
in Figure~\ref{fig4}), we obtained $L_3/L_{12} \simeq 0.52 \pm 0.05$,
so that we found that HT~Vir~A was about two times fainter than the
contact binary at its light maxima. 
In fact, we derived the ratio of the luminosities
to be about $0.48 \pm 0.03$ for phases around
0.25 and about $0.56 \pm 0.03$ 
for phases around 0.75; this may reflect either an
asymmetry in the light maxima of HT~Vir~B or systematic errors
in our estimates of $L_3/L_{12}$ for the mutually oppositely-orientated 
peaks in the broadening function.
After subtraction of the scaled peak of the third component, we
obtained an excellent radial-velocity orbit for HT~Vir~B (as
described in Section~\ref{sec2}).

We discovered that HT~Vir~A is a relatively close, single-lined 
(SB1) binary so that the HT~Vir system is a quadruple one.
The data for this star showed a scatter of about 20 km~s$^{-1}$ which we
analyzed for periodicity. Solution of the radial-velocity data,
using the program of \citet{mor75}, with a preliminary 
orbital period of 32.45 days, leads to an eccentric orbit with parameters
listed in Table~\ref{tab4}. 
We show the solution for HT~Vir~A in the last panel
of Figure~\ref{fig5}.

\section{SUMMARY}

The paper brings radial-velocity data and orbital solutions 
for the fourth group of ten close binary systems 
that we observed at the David Dunlap Observatory. 
Only 44~Boo and SW~Lyn have been observed spectroscopically before;
for both, we provide much improved spectroscopic orbits. 
All systems are double-lined 
(SB2) binaries with visible spectral lines of
both components. We give the values 
of $(M_1+M_2) \sin^3i = 1.0385 \times 10^{-7}\,(K_1+K_2)^3\,
P({\rm day})\,M_\odot$ in Table~\ref{tab2}. 
As in the previous papers of this series,
we have not been able to converted them into the sums of masses 
because in most cases the 
inclination angle is either unknown or not trustworthy. 
We note that the photometric discoveries of the
Hipparcos project (FI~Boo, V2150~Cyg, EX Leo, V2377 Oph) 
tend to have small values of $(M_1+M_2) \sin^3i$, in accord
with their small photometric amplitudes, both features
apparently due to low orbital inclinations. But we also found that
some of the newly-discovered 
systems may have small photometric amplitudes because
of the dilution of light in a triple system; the perfect example
is V899~Her. By discovering low photometric amplitude
binary systems, the Hipparcos mission has therefore contributed
substantially to rectifying the statistics of bright contact
binaries; this statistics had been known to be skewed toward
large amplitude variables \citep{rk94,ogle2}.

Five system in the current group are members of close 
visual and/or spectroscopic triple-lined systems 
(44~Boo, V899~Her, VZ~Lib, SW~Lyn, HT~Vir); among them
three and possibly four companions are themselves close binaries. 
We have been able to provide good data for these systems mostly
because of the superior capabilities of the linear SVD 
broadening-function approach \citep{ruc99} in resolving 
individual components in the triple-lined spectra.

\acknowledgements
The authors would like to thank Jim Thomson for help with
observations, Dr.\ Frank Fekel for providing us a list of 
early-type standard stars and Dr.\ Russ Robb for correspondence
about GSC~8--324. We thank the referee for a careful and constructive
review.

The visit of WO to the David Dunlap Observatory was possible thanks
to support from Professor Bohdan Paczynski, through
the Polish Astronomical Foundation.
Support from the Natural Sciences and Engineering Council of Canada
to SMR is acknowledged with gratitude. 

The research has made use of the SIMBAD database, operated at the CDS, 
Strasbourg, France and accessible through the Canadian 
Astronomy Data Centre, which is operated by the Herzberg Institute of 
Astrophysics, National Research Council of Canada.

\bigskip

\noindent
Errata to Paper II: 

\noindent
The residuals $\Delta V$ listed in Table~1 for SV~Equ
are incorrect and do not agree with Figure~1 and the quoted value of the
error per single observations, as given in Table~2. The correct residuals
can be computed from the original data and the spectroscopic elements.

\clearpage

\noindent
Captions to figures:

\bigskip

\figcaption[fig1_30w.ps] {\label{fig1}
Radial velocities of the systems 44~Boo~B, FI~Boo, V2150~Cyg and V899~Her~B
are plotted in individual panels versus orbital phases. The lines 
give the respective circular-orbit (sine-curve) fits to the radial 
velocities.
44~Boo~B and FI~Boo are W-type systems while V2150~Cyg and V899~Her~B
are A-type systems. Short marks in the lower 
parts of the panels show phases of available 
observations which were not used in the solutions because of the 
blending of lines. Open symbols in this and the next two figures
indicate observations given half-weights in the solutions. All
panels have the same vertical scales.
}

\figcaption[fig2_30w.ps] {\label{fig2}
Radial velocities of the systems EX~Leo, VZ~Lib~A, SW~Lyn~A and V2377~Oph.
Only V2377~Oph is a W-type system. SW~Lyn~A is most probably a short-period
Algol system and EX~Leo and VZ~Lib~A are contact systems of the A-type. 
}

\figcaption[fig3_30w.ps] {\label{fig3}
Radial velocities of the systems GSC~8-324 and HT~Vir~B.
The first is a very close, but detached or semi-detached system, while
HT~Vir~B is a contact A-type system with a surprisingly large mass-ratio.
}

\figcaption[fig4_30w.ps] {\label{fig4}
The broadening functions for five close binary systems with 
spectroscopic companions. For HT~Vir, 
we show (1) the full broadening function with the signature of
a sharp-line template shifted for clarity by 100 km~$^{-1}$
(dotted line in the right upper panel, shifted to right for
clarity) and (2) the same function
with the sharp-line component subtracted, leaving only the
signature of the contact binary HT~Vir~B (right lower panel). In each
panel the number gives the orbital phase selected to be close to
the orbital quadrature.
}

\figcaption[fig5_30w.ps] {\label{fig5}
The radial-velocity variations of the spectroscopic companions
of V899~Her~A, VZ~Lib~B, SW~Lyn~B and HT~Vir~A.  
For HT~Vir~A we show the radial-velocity data phased with period
of 32.45 days together with our preliminary single-line
orbital solution. The vertical scale in all panels has 
the same span of 60 km~s$^{-1}$, 
but the time axis is different for each panel.
}

\clearpage                    

\begin{table}                 
\dummytable \label{tab1}      
\end{table}
 
\begin{table}                 
\dummytable \label{tab2}      
\end{table}

\begin{table}                 
\dummytable \label{tab3}      
\end{table}

\begin{table}                 
\dummytable \label{tab4}      
\end{table}

\end{document}


\begin{deluxetable}{lclrrrccc}

\tabletypesize{\scriptsize}

\pagestyle{empty}
\tablecolumns{9}

\tablewidth{0pt}
\tablenum{2}
\tablecaption{Spectroscopic orbital elements of the fourth
ten of close binary systems}
\tablehead{
   \colhead{Name} &                
   \colhead{Type} &                
   \colhead{Other names} &         
   \colhead{$V_0$} &            
   \colhead{K$_1$} &               
   \colhead{$\epsilon_1$} &        
   \colhead{T$_0$ -- 2,400,000} &  
   \colhead{P (days)} &            
   \colhead{$q$}          \\       
   \colhead{}     &                
   \colhead{Sp.~type}    &         
   \colhead{}      &               
   \colhead{} &                    
   \colhead{K$_2$} &               
   \colhead{$\epsilon_2$} &        
   \colhead{(O--C), [E]} &         
   \colhead{$(M_1+M_2) \sin^3i$} &    
   \colhead{}                      
}
\startdata

44~Boo~B\tablenotemark{a} 
       &  EW/W & HD~133640B & $-17.89$ (0.40) & 231.31 (0.65) &
 5.50  & 50,944.6878(2)  & 0.267818    & 0.487 (6) \phn    \\
       & (K2~V) &  HIP~73695  &               & 112.70 (0.46) &
 3.29  & $+0.0002$ [8]   & 1.132 (11)  &               \\[1mm]

FI~Boo\tablenotemark{b}  
       &  EW/W & HD~234224  & $-30.55$ (0.72) & 148.65 (1.10) &
 5.17  & 51,718.3951 (7) & 0.389998    & 0.372 (21)    \\
       & G3~V &  HIP~75203 &                & 55.27 (0.90)  &
 4.52  & $+0.0926$ [8,252]& 0.343 (10)  &               \\[1mm]

V2150~Cyg &  EW/A & HD~202924  & $-12.82$ (0.45) & 125.43 (0.55) &
 3.18  & 51,386.3434 (7) & 0.591856    & 0.802 (6) \phn   \\
       & A6~V &  HIP~105162 &               & 156.40 (0.66) &
 5.69  & $+0.0195$ [4,876]& 1.376 (18)  &              \\[1mm]

V899~Her~B\tablenotemark{a}
                 & EW/A & HD~149684  & $-16.84$ (1.30) & 135.97 (2.04) &
 14.9  & 51,533.8405 (11)& 0.421173    & 0.566 (18)   \\
       & F5 comp. &  HIP~81191 &                & 240.37 (2.12) &
 14.5  & $-0.0046$ [7,203]& 2.331 (77)  &              \\[1mm]

EX~Leo &  EW/A & BD~17$^\circ$2269& $-11.05$ (1.10) & 51.28 (1.07) &
 3.23  & 51,615.6025 (13)& 0.408604    & 0.199 (36)   \\
       & F6~V &  HIP~52580  &               & 257.98 (1.85) &
 20.1 & $-0.0110$ [7,625]& 1.255 (36)  &              \\[1mm]

VZ~Lib~A\tablenotemark{a}
            &  EW/A &                 & $-31.11$ (2.30) & 68.54 (3.84) &
 18.5 & 51,091.4297 (15)& 0.358263    & 0.237 (68)   \\
      & G0 comp. &  HIP~76050 &                 & 289.25 (4.55) &
 27.7 & $-0.0814$ [17,593]& 1.704 (120)  &              \\[1mm]

SW~Lyn~A\tablenotemark{a,c} 
          &  EB & HD~67008         & $+32.39$ (1.35) & 116.73 (1.65) &
 11.2 & 51,400.1795 (25)& 0.644066     & 0.524 (28)   \\
       & (F2~V) &  HIP~39771 &                & 222.75 (3.20) &
 19.1 & $-0.0062$ [801]& 2.617 (112)   &              \\[1mm]

V2377~Oph &  EW/W & HD~159356  & $-25.79$ (0.38) & 159.64 (0.70) &
 3.18 & 51,720.3294 (5)& 0.425401      & 0.395 (12)   \\
       & G0/1~V &  HIP~85944  &               & 62.99 (0.62) &
 3.16 & $+0.0128$ [7,570]& 0.487 (9)   &              \\[1mm]

Anon Psc &  EB? EA? & GSC~8--324\tablenotemark{d}
                                   & $-27.85$ (0.73) & 131.62 (0.55) &
 4.44 & 51,794.1921 (4)& 0.308550      & 0.702 (14)   \\
       & (K4/5~V) &            &                & 187.55 (1.80) &
 11.6 & $-0.0188$ [1,110]& 1.042 (23)   &              \\[1mm]

HT~Vir~B\tablenotemark{a} 
                &  EW/A & HD~119931 & $-23.38$ (0.68) & 169.39 (1.00) &
 9.32 & 51,068.7101 (6)& 0.407670      &  0.812 (8) \phn  \\
       & F8~V comp. &  HIP~67186 &                & 208.54 (1.00) &
 9.45 & $+0.0382$ [17,230]& 2.285 (38)   &              \\

\tablenotetext{a}{Spectroscopically triple system.}
\tablenotetext{b}{FI~Boo: The period is twice the value given in the
   Hipparcos catalogue where the time of
   maximum light is given as the initial epoch. The large O--C reflects
   this difference.}
\tablenotetext{c}{The solution for SW~Lyn~A is based on both observational
   seasons. See the text for discussion concerning
   the possible influence of the motion around the common center of
   mass in the triple system.}
\tablenotetext{d}{The Guide Star Catalogue
   number is written in an abbreviated form in the
   table. The correct number is: GSC~00008--00324.}
\tablecomments{The spectral types given in column two
are new except those in parantheses which are either taken from the
literature or estimated from color indices. The convention of naming the
binary components is that the subscript 1 designates
the component which is eclipsed at the
deeper minimum and is therefore the hotter one. 
The standard errors of the circular 
solutions in the table are expressed in units of last decimal places 
quoted; they are given in parantheses after each value. 
For example, the last table entry for the mass ratio 
$q$, 0.812~(8), should be interpretted as $q = 0.812 \pm 0.008$;
the mass ratio is defined to be always $q \le 1$.
The center-of-mass velocities ($V_0$), the 
velocity amplitudes (K$_i$) and the standard unit-weight 
errors of the solutions ($\epsilon$) are expressed 
in km~s$^{-1}$. The spectroscopically determined 
moments of primary minima are given by $T_0$; the corresponding
(O--C) deviations (in days) have been calculated from the most 
recent available ephemerides, as given in the text,
using the assumed periods and the number of epochs given by [E].
The values of $(M_1+M_2)\sin^3i$ are in the solar mass units.
}
\enddata
\end{deluxetable}